\documentclass[12pt,preprint]{aastex}

\newcommand{\myemail}{tsuchiya@icrr.u-tokyo.ac.jp,enomoto@icrr.u-tokyo.ac.jp}

\slugcomment{To appear in ApJ Letters.}

\shorttitle{Sub-TeV Gamma-rays from the Galactic Center}
\shortauthors{Tsuchiya, Enomoto et al.}

\begin{document}

\title{Detection of sub-TeV Gamma-rays from
the Galactic Center Direction by CANGAROO-II}


\author{
K.~Tsuchiya\altaffilmark{1}, 
R.~Enomoto\altaffilmark{1}, 
L.T.~Ksenofontov\altaffilmark{1}, 
M.~Mori\altaffilmark{1}, 
T.~Naito\altaffilmark{2},
A.~Asahara\altaffilmark{3}, 
G.V.~Bicknell\altaffilmark{4}, 
R.W.~Clay\altaffilmark{5}, 
Y.~Doi\altaffilmark{6},   
P.G.~Edwards\altaffilmark{7}, 
S.~Gunji\altaffilmark{6}, 
S.~Hara\altaffilmark{1}, 
T.~Hara\altaffilmark{2}, 
T.~Hattori\altaffilmark{8}, 
Sei.~Hayashi\altaffilmark{9},  
C.~Itoh\altaffilmark{10}, 
S.~Kabuki\altaffilmark{1}, 
F.~Kajino\altaffilmark{9}, 
H.~Katagiri\altaffilmark{1}, 
A.~Kawachi\altaffilmark{1}, 
T.~Kifune\altaffilmark{11},
H.~Kubo\altaffilmark{3}, 
T.~Kurihara\altaffilmark{8},  
R.~Kurosaka\altaffilmark{1}, 
J.~Kushida\altaffilmark{8} 
Y.~Matsubara\altaffilmark{12}, 
Y.~Miyashita\altaffilmark{8},
Y.~Mizumoto\altaffilmark{13},
H.~Moro\altaffilmark{8},  
H.~Muraishi\altaffilmark{14},
Y.~Muraki\altaffilmark{12},
T.~Nakase\altaffilmark{8}, 
D.~Nishida\altaffilmark{3}, 
K.~Nishijima\altaffilmark{8}, 
M.~Ohishi\altaffilmark{1}, 
K.~Okumura\altaffilmark{1},  
J.R.~Patterson\altaffilmark{5}, 
R.J.~Protheroe\altaffilmark{5}, 
N.~Sakamoto\altaffilmark{6},   
K.~Sakurazawa\altaffilmark{15},
D.L.~Swaby\altaffilmark{5}, 
T.~Tanimori\altaffilmark{3}, 
H.~Tanimura\altaffilmark{3}, 
G.~Thornton\altaffilmark{5}, 
F.~Tokanai\altaffilmark{6}, 
T.~Uchida\altaffilmark{1},
S.~Watanabe\altaffilmark{3},  
T.~Yamaoka\altaffilmark{9},  
S.~Yanagita\altaffilmark{16}, 
T.~Yoshida\altaffilmark{16}, 
T.~Yoshikoshi\altaffilmark{17}}

\altaffiltext{1}
{ICRR, Univ.\ of Tokyo, Kashiwa, Chiba 277-8582, Japan; \myemail}
\altaffiltext{2}
{Fac.\ of Management Information, Yamanashi Gakuin Univ., 
Yamanashi 400-8575, Japan}
\altaffiltext{3}
{Dept.\ of Phys., Grad.\ School of Sci., Kyoto Univ., 
Kyoto 606-8502, Japan}
\altaffiltext{4}
%{Research School of Astronomy and Astrophysics, Australian National 
{RSAA, Australian National Univ., ACT 2611, Australia}
\altaffiltext{5}
%{Dept. of Phys. and Mathematical Phys., Univ. of
{Dept.\ of Physics, Univ.\ of 
Adelaide, SA 5005, Australia}
\altaffiltext{6}
{Dept.\ of Phys., Yamagata Univ., Yamagata 990-8560, Japan}
\altaffiltext{7}
{ISAS/JAXA, Sagamihara, Kanagawa 229-8510, Japan} 
\altaffiltext{8}
{Dept.\ of Phys., Tokai Univ., Kanagawa 259-1292, Japan}
\altaffiltext{9}
{Dept.\ of Phys., Konan Univ., Hyogo 658-8501, Japan}
\altaffiltext{10}
{Ibaraki Prefectural Univ.\ of Health Sci., Ibaraki 300-0394, Japan}
\altaffiltext{11}
{Fac.\ of Engineering, Shinshu Univ., Nagano 480-8553, Japan}
\altaffiltext{12}
{STE Lab.,  Nagoya Univ.,  Aichi 464-8602, Japan}
\altaffiltext{13}
{NAOJ, Mitaka, Tokyo 181-8588, Japan}
\altaffiltext{14}
{School of Allied Health Sci., Kitasato Univ., Kanagawa 228-8555, Japan}
\altaffiltext{15}
{Dept.\ of Phys., TIT,  Tokyo 152-8551, Japan}
\altaffiltext{16}
{Fac.\ of Sci., Ibaraki Univ., Ibaraki 310-8512, Japan}
\altaffiltext{17}
{Dept.\ of Phys., Osaka City Univ., Osaka 558-8585, Japan}


\begin{abstract}

We have detected sub-TeV gamma-ray emission from the direction of
the Galactic Center (GC) 
using the CANGAROO-II Imaging Atmospheric Cherenkov Telescope
(IACT).
We detected a statistically significant excess
at energies greater than 250\,GeV.
The flux was one order of magnitude lower than that of Crab
at 1~TeV
with a soft spectrum $\propto E^{-4.6\pm0.5}$.
The signal centroid is consistent with the GC direction and
the observed profile is consistent with a point-like source.
Our data suggests that
the GeV source 3EG~J1746$-$2851 is identical with this TeV source and 
we study the combined spectra to determine the possible origin
of the gamma-ray emission.
We also obtain an upper
limit on the cold dark-matter density in the Galactic halo.
\end{abstract}




\section{Introduction}

Many observations have been made of
the Galactic Center (GC) region over a wide range of wavelengths
\citep[e.g.,][]{pauls76,sofue86,pedlar89,koyama96,buckley97,
purcell97,mh98,maeda02}.
At the dynamical center of the Galaxy is the source Sgr~A$^*$,
containing a 4$\times 10^6$\,M$_\sun$ black hole, which is
a strong, variable, radio source 
\citep[e.g.,][]{zhao03}
and 
a weak, but occasionally flaring, infra-red \citep{genzel}
and X-ray \citep{baganoff} source.
Surrounding Sgr~A$^*$ are 
the (presumed) supernova remnant Sgr~A~East, and 
a number of filamentary radio structures,
including one known as the Arc, extending over tens of parsecs.
One filamentary structure was recently found to be a non-thermal X-ray 
source \citep{sakano}. 
The radio emission is interpreted as synchrotron radiation
from high-energy electrons in the ambient magnetic fields.
The observed magnetic fields are as high as 
several
mG, although whether
they are distributed over the whole GC region or are localized is not
known \citep{morris}. 

The bright EGRET source 3EG~J1746$-$2851 is listed as unidentified
in the third EGRET catalog \citep{hartman99}, but as the position,
given in 3EG as
($l=0.11, b=-0.04$) with an 95\% confidence contour radius of 0.13 degrees,
is consistent with the GC, it has been considered as
the gamma-ray counterpart to GC region \citep{mh98}. 
Three different source scenarios were suggested by
\cite{mh98}: emission from one or more pulsars, inverse Compton radiation
from relativistic electrons in the Arc, or Cold Dark Matter (CDM)
annihilation.
N-body simulations suggest a significant
enhancement of the CDM density around galaxy cores \citep{navarro96}.
Better constraints on CDM can be obtained
by high-energy gamma-ray observations of the GC region.
More recently, 3EG~J1746$-$2851 has been associated with Sgr~A~East,
with the gamma-ray emission arising from the decay of neutral pions
produced by high-energy protons,
accelerated in the remnant, interacting with the ambient matter \citep{fm03}.
Time variability was reported, 
although its amplitude is not large \citep{no03}.  

Spectra derived from the EGRET data 
\citep[][shown in Fig. 3]{mh98,hartman99} are not well fit
by a single power-law: the spectra are quite hard at low energies but
flatten, or possibly turn over, at high energies.
Simple extrapolations of the high energy spectra, however, 
suggest a potentially detectable TeV flux.

We have observed the GC region with the
CANGAROO-II Imaging Atmospheric Cherenkov Telescope (IACT).
The 10\,m diameter CANGAROO-II telescope \citep{kawachi}
detects gamma-rays above several hundred GeV
by detecting optical
Cherenkov radiation generated by relativistic secondary particles in
the cascades produced when high-energy gamma-rays (and background cosmic rays)
interact with the Earth's upper atmosphere. 
The optical images provide information on the 
direction and energy of gamma-ray events.
The telescope is located near Woomera, South Australia 
(136$^\circ$47$'$\,E, 31$^\circ$06$'$\,S) and thus the GC
culminates within $\sim$2$^\circ$ of the zenith, enabling 
observations with a lower energy threshold.
Observations with the Whipple IACT from Mt Hopkins (31$^\circ$58$'$\,N)
yielded an 2\,$\sigma$ upper limit of 
$0.45\times 10^{-11}$\,cm$^{-2}$\,s$^{-1}$ at 2\,TeV \citep{buckley97}.

The angular resolution of CANGAROO-II is 0.24$^\circ$
(36\,pc at a distance of 8.5\,kpc) with an energy
threshold of 400\,GeV in the case of a Crab-like energy spectrum 
($\propto E^{-2.5}$). 
For the softer spectrum, $\propto E^{-4.6}$, we find (see \S\,4),
the angular resolution becomes 0.32$^\circ$ (47\,pc at 8.5\,kpc) and 
the energy threshold is lowered to 250\,GeV.
With this angular resolution we cannot resolve 
Sgr~A$^*$ ($\sim$1\,pc in size) and the elliptical Sgr~A~East 
(major axis of 10.5\,pc), 
which surrounds Sgr A$^*$ \citep{yusef,maeda02}. 
The location of the EGRET source 3EG~J1746$-$2851 
is also too close to the GC for us to resolve it.

\section{Observations}

The observations were carried out during 2001 July 12--24 (13 nights), and 
between 2002 July~4 and August~11 (20 nights). 
The telescope tracked Sgr~A$^*$ 
($\alpha =$266.42$^\circ$, $\delta =-$29.01$^\circ$, J2000 coordinates).
The field of view (FOV) of the camera is 2.76$^\circ \times$2.76$^\circ$.
The brightest star in the FOV (SAO 185755) has a visual magnitude of 4.7. 
Most stars in this region are reddened, and the camera 
is sensitive to UV light, however this region is brighter 
than our typical on-source FOV. 
As a result, we used a higher trigger threshold,
requiring 4 triggered pixels rather than the usual 3 \citep{cito},
which reduced the trigger rate from 17 to 6\,Hz.
Lights from the detention center located several kilometers from
the telescope had little effect on these observations as the telescope
was pointed close to the zenith \citep{cito}.
This was checked from the azimuthal angle dependence of the shower rate
(see \S\,3).
Each night was divided into two or three periods, i.e., ON--OFF,
OFF--ON--OFF, or OFF--ON observations. ON-source observations were timed
to contain the meridian passage of the target, 
as was done by %the same procedure with 
\cite{nature}.
In total, 7300 min.\ of ON- and 7100 min.\ of OFF-source data
were obtained.

\section{Analysis}

First, `cleaning' cuts on camera images were applied,
requiring (0.115$^\circ$-square) pixel pulse-heights of greater than
3.3 photoelectrons, and Cherenkov photon arrival times within $\pm$40~ns
of the median arrival time.
Clusters of at least five adjacent triggered pixels 
were required in each event
(in contrast to our usual acceptance of clusters of at least four pixels)
to minimize the effects of the bright star field.
After these pre-selection cuts, the shower rate was stable on a
run-to-run basis for observations in the same year.
Events due to background light were reduced by 99.8\%
at this level. 
A difference of $18\pm14$\% in the average 
shower rates for 2001 and 2002 was apparent at this stage.
We interpret this as being due to a deterioration of the mirror reflectivity
and/or camera sensitivity, of $-10.2\pm4.6$\%. This was
taken into account in further analysis and included in 
the final systematic errors in Table \ref{table1}.
The systematic difference of the run-by-run acceptance within the
same year is 
expected to be less than 8.4\%.
The ON/OFF shower rate difference was ($3\pm9$)\%.
By examining the event rates within each run 
we were able to reject periods affected by cloud, 
dew forming on the mirrors,
instrumental abnormalities,
etc.
Only data taken at elevation angles greater than 60$^{\circ}$ were accepted.
After these cuts, 4000 min.\ of ON- and 3400 min.\ of OFF-source
data survived.

Trigger rates for each pixel per 700\,$\mu$s were monitored by a 
scaler circuit in real-time and recorded each second.
These data were used to exclude `hot' pixels (generally due to 
the passage of a star through the FOV of a pixel) in off-line
analysis.
Hillas parameters were then calculated to discriminate gamma-rays
from cosmic-rays based on the image shape and orientation  \citep{hillas}.
Further we masked 
a small number of 
pixels which showed deformed ADC spectra,
possibly due to a hardware fault.
Discrimination of the cosmic-ray background from gamma-rays
was carried out using the likelihood method of \cite{enomotoapp}.

Crab nebula data were analyzed with the same code, with
the derived flux and morphology consistent with 
the previous measurements and the point-spread function, respectively.
The distributions of the Hillas parameters for the excess events
were checked and found to be consistent with Monte Carlo
simulations for gamma rays. The OFF-source data
were compared with the Monte Carlo simulations of protons,
and found to be consistent.

\section{Results}

\begin{figure}
\plotone{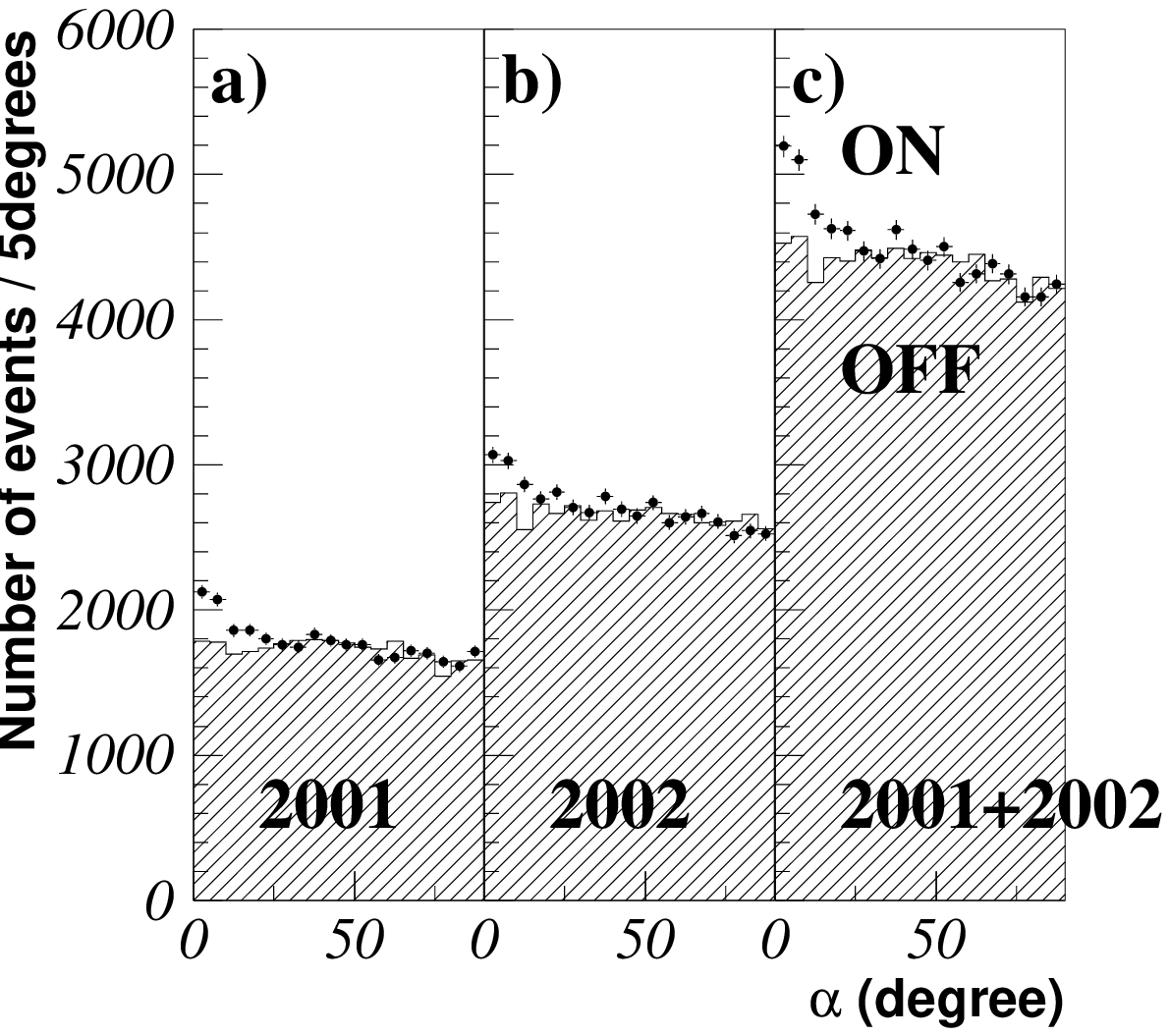}
\caption{Distributions of $\alpha$ (image orientation angle).
From left to right, the
a) 2001 data, b) 2002 data, and c) combined data,
are shown.
The points with error bars show the ON-source data 
and the hatched histograms are the normalized OFF-source data. 
}  
\label{fig1}
\end{figure}

The resulting distributions of the image orientation angle, $\alpha$, 
are shown in Fig.~\ref{fig1} for both years and for the combined data-set.
The normalizations between the ON- and
OFF-distributions were carried out using data with
$\alpha>30^\circ$.
The numbers of excess events ($\alpha <15^\circ$) 
were 800$\pm$100 (in an observation time of 1400 min.), 860$\pm$140
(2600 min.), and 1660$\pm$170 (4000 min.) in 2001, 2002, 
and the combined data, respectively, where the quated errors include only 
statistical ones.
A difference in the excess rates for 2001 and 2002 is apparent,
with the (2001/2002) ratio being 1.60$\pm$0.34. 
We note that a flare was observed at 1.3\,mm from Sgr~A$^*$ in July 2001
\citep{yuan}.
However the shower rate (effectively energy threshold)
difference described above could affect this ration at the 20\% level. 
We are, thus, unable to infer that the TeV emission is from a variable source. 
Nightly signal rates were also checked during both years, 
even though these have poorer statistics.
The largest deviation occurred on 2002 Aug 11, UT, 
with a rate $1.8\pm0.6$ times larger than the average for that year, 
again not statistically significant. 

\begin{figure}
\plotone{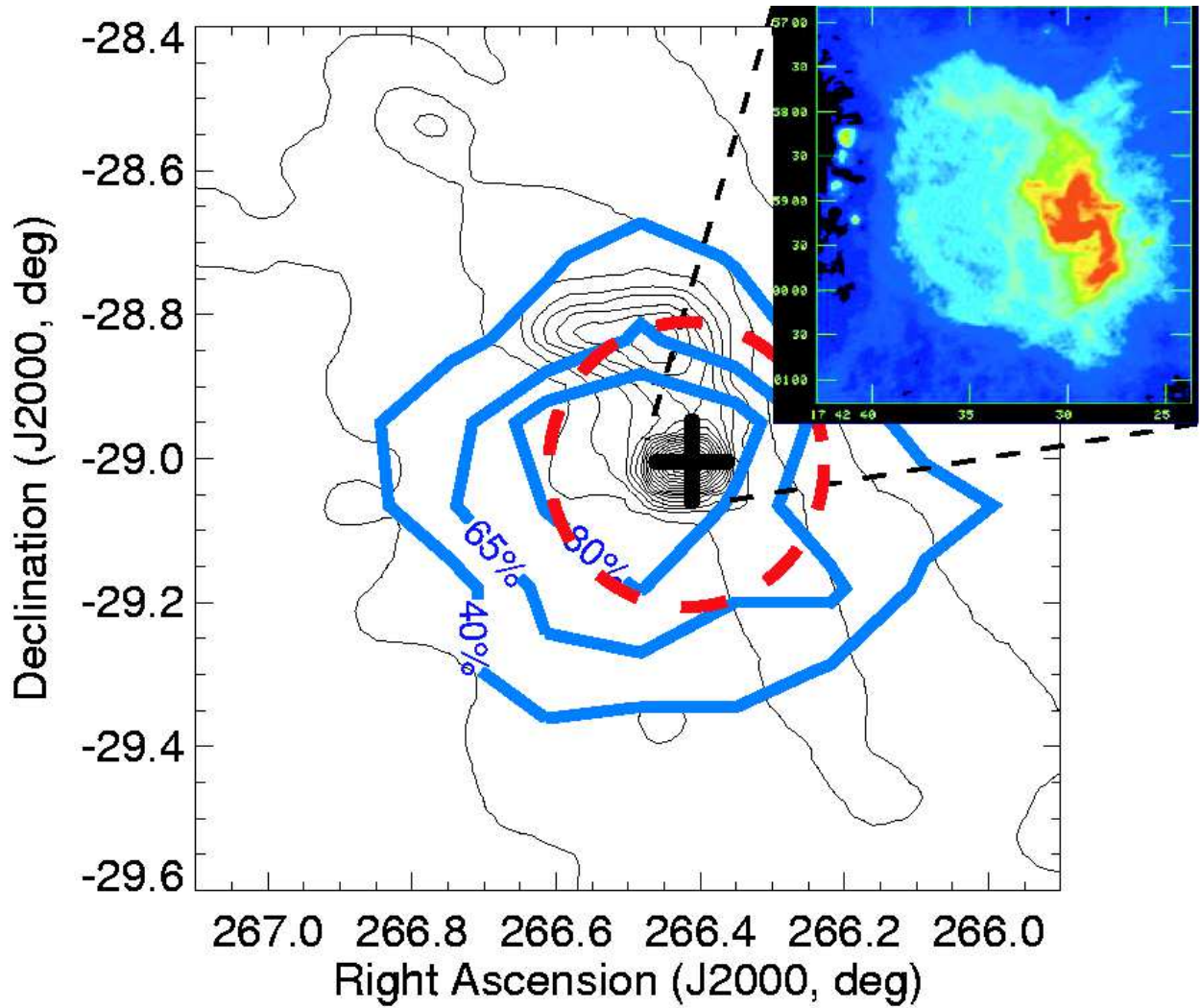}
\caption{ 
The ``significance map" obtained by the CANGAROO-II telescope 
is shown by the blue contours. 
The thin contours are a 12$\mu$ IRAS image.
The position of Sgr~A$^*$ (the telescope tracking center) 
is given by the cross. 
The inset is a 5\,GHz VLA image showing Sgr~A$^*$ and Sgr~A~East
\citep{yusef}.
The uncertainty in the position for 3EG~J1746$-$2851
analysed by \cite{mh98} is indicated by
the orange dashed contour.
% taken from \cite{mh98}.
\label{fig2}}
\end{figure}

Our Monte Carlo simulations predict, for a point-source,
that gamma-ray events with
$\alpha<15^\circ$ should constitute 73.5\% of those with $\alpha<30^\circ$.
The experimental data yielded 
$80.8\pm 6.7$\%, consistent with the point-source assumption.
As a further check on the spatial distribution of the signal, 
we derived the ``significance map'', as shown by the thick contours in 
Fig.~\ref{fig2}.
The contours were
calculated from the distribution of the detection significance
determined at each location from the difference in the $\alpha$ plots
(ON- minus OFF-source histogram) divided by the statistical errors.
The centroid is consistent with 
3EG\,J1746$-$2851, within
our possible systematic uncertainty of 0.1$^\circ$,
and so we identify the GeV source as the likely
origin of the TeV emission.
Our angular resolution was estimated to be 0.32$^\circ$, 
slightly larger than the radius of the 65\% contour, consistent
with the point-source assumption. 
The acceptance of the CANGAROO-II telescope is a smoothly
decreasing function with an offset from the tracking center, falling to
50\% at a 0.9$^\circ$ offset \citep[see][for details]{cito}. 

After correcting for this acceptance, the differential fluxes 
listed in Table~\ref{table1} were derived.
As both statistical and systematic errors are included,
the energy bins overlap somewhat, particularly at low energies. 
The systematic uncertainty for the energy determination
($\sim$20\%), which is an overall factor,
dominates the errors in the energies.
The Spectral Energy Distribution (SED) is plotted in Fig. \ref{fig3}
together with the EGRET data \citep{mh98,hartman99}.
The cross-hatched area indicates the CANGAROO-II data reported here.
The derived spectrum has a power-law index
of $-4.6\pm0.5$, much softer than that of the Crab nebula ($-$2.5).
When considered together with the EGRET data it is clear the
flux falls off steeply in the TeV region.

Various checks on the signal level and position were carried out, by
varying thresholds, clustering cuts, Hillas parameter values, etc:
these yielded consistent fluxes within the systematic
errors given in Table \ref{table1}. 

\section{Discussion}
Due to the complex structure of the GC region different scenarios for
the origin of the EGRET gamma-ray flux have been considered
\citep[see, e.g.,][and references therein]{mh98}.
As mentioned in \S1,
\cite{fm03} have recently attributed the GeV emission to 
$\pi^0$ decay resulting from
high-energy protons interacting with the ambient matter in
Sgr~A East.
They argued that due to synchrotron cooling in the high average
magnetic field, primary (accelerated) and secondary leptons would
have a steep spectrum, and their contribution to the high energy 
gamma-ray flux via inverse Compton emission and bremsstrahlung would be 
small. 
The leptonic contribution to the TeV gamma-ray flux would be even smaller.
Indeed, the synchrotron cooling
time can be estimated as $\tau_s=16\times B_{mG}^{-2}
E_{e,TeV}^{-1}$~yr. Thus for leptons of a few TeV, which might contribute to
the sub-TeV gamma-ray flux, the cooling time
in a mG magnetic field is even less than those
considered by \cite{fm03}.  Another factor which will reduce the amount
of secondary TeV leptons, the key products from {\it pp} interactions,  is 
the presence of a cut-off in the accelerated proton spectrum at a few 
TeV, suggested by this observation (see below). 
In the case of Sgr~A$^*$, an even higher magnetic field
is expected, which would lead to the same conclusion.

\begin{figure}
\plotone{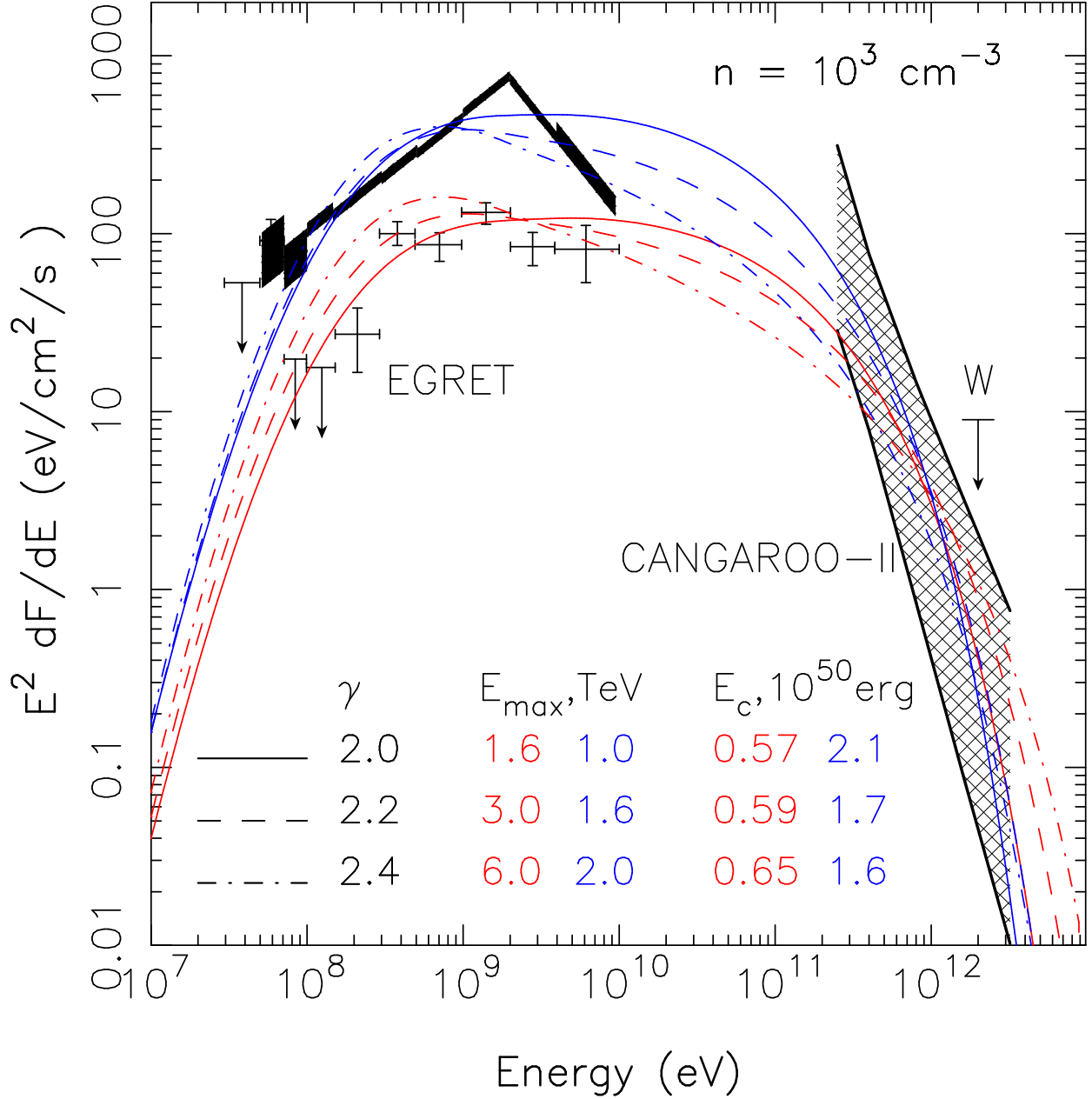}
\caption{ 
Spectral energy distribution of the GC region.
The cross-hatched area is the 1\,$\sigma$ allowed region
for the TeV observations in the energy range of Table~\ref{table1}. 
Here the energy uncertainties in Table~\ref{table1} were
assumed to be correlated bin by bin.
The arrow (W) is the Whipple 2\,$\sigma$ upper limit at 2~TeV
\citep{buckley97}. 
The two analyses of the EGRET data are shown by the 
black hatched region \citep{mh98} and the crosses \citep{hartman99}.
The lines are estimations for $\pi^0$ gamma-rays,
the details of which are given in the body of the figure and in the text.
\label{fig3}}
\end{figure}

In Fig.~\ref{fig3} we reproduce the $\pi^0$ decay emission to fit both
the EGRET data and ours. The input proton spectrum was assumed to be
proportional to $E^{-\gamma}e^{-E/E_{max}}$. The number density of
the ambient gas ($n$) was taken to be $\sim 10^3$ cm$^{-3}$
\citep{maeda02}, however this affects only the estimation of the
total proton cosmic-ray energy, as $E_c\propto n^{-1}$. 
The various lines shown in Fig.~\ref{fig3} are for differing values
of $\gamma$ and $E_{max}$. The GeV spectrum given in the 3EG
catalog \citep{hartman99} differs from that of the dedicated analysis
of \cite{mh98}, however, in both cases
the EGRET and CANGAROO-II data can
be relatively smoothly connected, with a cutoff energy of 1--3\,TeV. 
The spectra for $\gamma=2.4$ yield slightly worse fits to the
data at the lowest energies.
The inferred total cosmic-ray energy greater than 1\,GeV (see Fig.~\ref{fig3})
corresponds to $\leq 10$\%
of a typical supernova energy, a quite plausible conversion efficiency.
The details, however, are dependent on the EGRET flux. In the
case of a lower gas density ($n$), the cosmic ray energy 
exceeds that of a single
supernova, and we should consider the possibility of 
other sources in the GC region, such as Sgr A$^*$ itself,
contributing to the gamma-ray flux. Our
data determine the maximum energy to which cosmic rays are accelerated 
in this region, which does not depend upon other physical parameters. 
For example, the range of values
is consistent with recent theoretical predictions for shock
acceleration in supernova remnants \citep{pz03}. 

Thus, identifying the TeV source with 3EG\,J1746$-$2851, and
associating these in turn with Sgr~A~East and/or Sgr A$^*$, we find that the 
gamma-ray fluxes and spectra can be naturally explained as arising
due to $\pi^0$ decay. 

We can, in addition, use our observations to 
derive upper limits to the CDM abundance in the GC region,
assuming the GeV and TeV emission is centered on Sgr~A$^*$ and
following the 
method of
\cite{enomoto253}.
The emission region is assumed to be a sphere with a radius
of 47\,pc.
The annihilation mode of $\chi \chi \rightarrow q\bar{q}$ 
was first considered.
The cross section ($\sigma$) multiplied by the
relative velocity ($v$) and the branching fraction ($B$)
is a free parameter that was normalized to $10^{-26} cm^3s^{-1}$.
The experimental data of $e^+e^-\rightarrow q\bar{q}\rightarrow \gamma X$
was used for the gamma-ray multiplicity \citep{lep}.
The derived 2\,$\sigma$ upper limits for
the CDM densities 
($\rho_{CDM}^{47pc}\sqrt{\sigma vB/10^{-26} cm^3s^{-1}}$)
are 9300, 7300, 5800, 5300, and 5800\,GeV/cm$^3$
for an assumed WIMP mass of 0.7, 1, 2, 4, and 6\,TeV, respectively. 
Here, although we took the EGRET flux as an upper limit, above 0.6\,TeV
it is the CANGAROO-II data that dominates these results.
\cite{navarro96} and subsequent analyses 
\citep{fukushige,moore,ghigna,jing,power} 
adopted a cusp structure, i.e.,
the density profile proportional to 
$r^{-\beta}(1+r/r_s)^{\beta-3}$, where $r_s$ is a free parameter
and $\alpha$ ranges from 1 to 1.5.
In the case of $\beta$=1.3, 
the upper limits of local densities 
($\rho_{CDM}^{\odot}\sqrt{\sigma vB/10^{-26} cm^3s^{-1}}$)
are of the order of several GeV/cm$^3$.
Higher upper limits are obtained when
the decay mode of $\chi \chi \rightarrow \gamma \gamma$ was assumed
with $\sigma vB_{\gamma \gamma}=10^{-29} cm^3s^{-1}$.

\acknowledgments
This work was supported by a Grant-in-Aid for Scientific Research by
the Japan Ministry of Education, Culture, Sports, Science and Technology, 
and by JSPS Research Fellowships.
This work was also supported by ARC Linkage Infrastructure Grant 
LE0238884  and Discovery Project Grant DP0345983.
We thank the Defense Support Center Woomera and BAE Systems.

\clearpage


\clearpage

\begin{table}   
\begin{center}
\caption{Differential fluxes.}
\label{table1}
\begin{tabular}{cl} 
\hline          
\hline               
Mean energy of bin [GeV] ~ ~ ~   &  Flux [ph/cm$^2$/s/TeV]\\
\hline
258$\pm$64   & (0.14$\pm$0.06)$\times$$10^{-8}$ \\
299$\pm$66   & (0.82$\pm$0.22)$\times$$10^{-9}$ \\
367$\pm$80   & (0.20$\pm$0.08)$\times$$10^{-9}$  \\
540$\pm$102  & (0.54$\pm$0.19)$\times$$10^{-10}$ \\
962$\pm$356  & (0.30$\pm$0.28)$\times$$10^{-11}$ \\
2454$\pm$653 & (0.73$\pm$1.19)$\times$$10^{-13}$\\
\hline
\hline
\end{tabular}
\end{center}
\end{table}

\end{document}